\documentclass{epl}

\usepackage{graphicx}
\usepackage{dcolumn}
\usepackage{bm}
\usepackage[latin1]{inputenc}
\usepackage{psfrag}
\usepackage{epic}
\usepackage{bm}
\usepackage{latexsym}
\usepackage{dsfont}
\usepackage{amsmath}

\newcommand{\trace}[1]{\mathrm{Tr}\, #1}

\newcommand{\expt}[1]{\langle #1\rangle}
\newcommand{\bra}[1]{\langle #1|}
\newcommand{\ket}[1]{|#1\rangle}

\newcommand{\transpose}{\ensuremath{\mathrm{T}}}
\newcommand{\expe}{\mathrm{e}}
\newcommand{\imi}{\mathrm{i}}

\begin{document}

\title{Thermalization through unitary evolution of pure states}
\shorttitle{Thermalization through unitary evolution etc.}

\author{S. O. Skrøvseth\thanks{E-mail: \email:stein.skrovseth@ntnu.no}}
\institute{%
  Department of Physics,
  Norwegian University of Science and Technology,
  N-7491 Trondheim, Norway
}

\date{\today}

\begin{abstract}
  The unitary time evolution of a critical quantum spin chain with an
  impurity is calculated,
  and the entanglement evolution is shown. Moreover, we show that the
  reduced density matrix of a part of the chain evolves such that the
  fidelity of its spectrum is very high with respect to a state in
  thermal   equilibrium. Hence, a thermal state occurs through
  unitary time evolution in a simple spin chain with impurity.
\end{abstract}

\maketitle

\section{Introduction}
Systems that are left alone to evolve in interaction with an
environment at a certain temperature, usually relax into a state in
thermal equilibrium with its environment, and in accordance with the
ergodic 
hypothesis, this has been shown experimentally for numerous examples
in classical 
systems. How this happens has been the subject of intense research for many
decades. The corresponding quantum
case is usually viewed as a stochastic prosess of "quantum jumps",
\cite{Feynman}
where only the transition probabilities are calculated from the
underlying unitary quantum dynamics (connected by the Fermi Golden
rule). 
The hypotheses behind this paper, however, is that thermalization
should generally occur in quantum systems describing a macrosopic
world, even when all its dynamics is described by pure unitary
evolution. 

Quantum spin chains have been a matter of keen investigation over the
latest years, and in particular their critical properties have been
investigated thoroughly 
since it was discovered that the non-classical correlations known as
entanglement are characteristic in quantum phase
transitions \cite{OsterlohSachdev}. This gives rise to the
study of entanglement in the chains, in particular at critical points,
as a part of the vast and expanding field of quantum information
science \cite{Nielsen&Chuang}. Moreover, it has been identified that
at critical points, conformal symmetry arises in a large class of
models, and the characteristics of conformal field theory can be used
to describe the universal properties of such systems
\cite{HolzheyVidalSOS}. For a certain class of models the
chain can be fermionized, which makes them more accessible to
analytic and numerical investigation \cite{SOSent}. Indeed, the
propagation of entanglement is complicated \cite{Amico04}, but can
be described by conformal field theory approaches \cite{CFTs}, and
scaling laws for the entanglement have been found \cite{Eisert06}

Any existent
entanglement measured by the concurrence (i.e. the entanglement of
formation of two spins) in the quantum spin chain at  
zero temperature disappears when the temperature of the system exceeds
a threshold temperature, see e.g. \cite{Brennen04}. However, the
entanglement entropy does not vanish, indeed conformal field theory
predicts that an infinitely long conformally invariant state in thermal
equilibrium at temperature $\beta$ has entropy proportional to the
temperature at high temperatures \cite{Calabrese04}. The entropy is
not an entanglement measure if the entire system is not in a pure
state, such as a thermal state. The dynamics resulting in
thermalization have also been discussed in other recent works
\cite{Thermals}.

\begin{figure}[htb]
  \centering\onefigure[scale=.75]{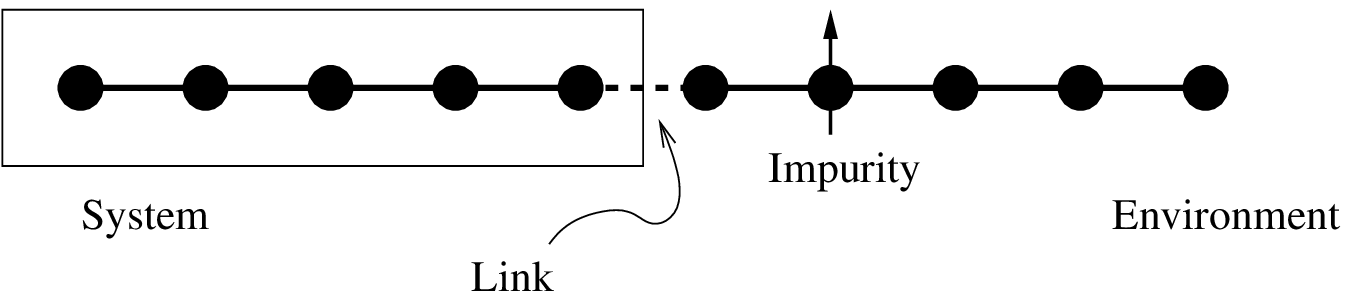}
  \caption{Conceptual sketch of the spin chain under
    consideration. Each
    filled circle is a spin site, and the links between them is
    indicated through lines. The location of the magnetic impurity is
    not fixed. The entire chain is in a pure state $\ket\psi$ while the
    subsystem is described by the reduced density matrix $\rho$. }
  \label{fig:sketch}
\end{figure}

In this paper we consider a quantum spin chain as sketched in
Fig. \ref{fig:sketch}, 
{\it i.e.} a system of $N$ spins with open boundary conditions. The chain
is initially in some arbitrary excited pure state $\ket\psi$ with
finite excitation energy per site, and evolves according to a unitary
time evolution. We trace
out some part of the system, which we denote the environment, to obtain the
reduced density matrix $\rho=\trace'\ket\psi\bra\psi$ of the
subsystem which consists of $N'$ spins. At some spin site
we introduce a magnetic impurity, that is an additional local magnetic
field. In the last section of this letter we 
also consider what happens when the link between the system and the
environment is cut. We will consider both the 
entropy of the state $\rho$, and the characteristics of $\rho$ as
related to a state in thermal equilibrium with its environment.

\section{Model}
We consider a critical Ising chain with open boundary conditions, and
with a magnetic impurity at site $\alpha$,
\begin{equation}
  \mathcal
  H=-\sum_{n=1}^N\left(\sigma_n^x\sigma_{n+1}^x+\sigma_n^z\right)+\delta\sigma_\alpha^z,\qquad\sigma_{N+1}^x=0.
  \label{eq:H}
\end{equation}
Here $\sigma^{(x,y,z)}_n$ are the Pauli spin matrices at site $n$.
$\delta\geq0$ is the strength of the extra magnetic field at the
impurity. If $\delta=0$ we recover the critical Ising model. The
impurity will destroy many of the symmetries in the 
system, and thereby the conservation laws, which ensures that the
initial state will evolve non-trivially.

We configure the system in the initial state
$\ket{\psi_0}=\ket{\psi(t=0)}$ which is not an eigenstate of the above
Hamiltonian. Rather, we choose it as an excited eigenstate of the
unperturbed Hamiltonian $\mathcal H_0$. Thus it will be ``almost'' an
eigenstate of the full Hamiltonian, and its time evolution will be
nontrivial.

To the Hamiltonian (\ref{eq:H}) we apply a Jordan-Wigner
transform, making it possible to map the model onto a string of
fermions \cite{JordanWigner}. This amounts to defining the fermionic
operators  
$\hat
c_n=\frac12\left(\bigotimes_{k=1}^{n-1}\sigma_k^z\right)\otimes(\sigma_n^x+\imi\sigma_n^y)$
and their adjoints, such that $\{\hat c_m,\hat
c_n^\dag\}=\delta_{mn}$. Furthermore we define the $2N$ Majorana fermions, 
which map two Majorana fermions onto one of the fermions defined by
$\hat c_n$,
\[\check\gamma_{2n-1}=\frac1{\imi\sqrt2}\left(\hat c_n-\hat
  c_n^\dag\right)\qquad
\check\gamma_{2n-1}=\frac1{\sqrt2}\left(\hat c_n+\hat
  c_n^\dag\right)\qquad
\{\check\gamma_i,\check\gamma_j\}=\delta_{ij}.\]
This diagonalizes the Hamiltonian in the sense that
$\mathcal H=\sum_{ij}C_{ij}\check\gamma_i\check\gamma_j$
where $C$ is an imaginary block diagonal matrix with $N$ antisymmetric
$2\times 2$ blocks.
Finally, we define the correlation matrix
$\Gamma_{ij}=\expt{[\check\gamma_i,\check\gamma_j]}$. From this one
can then trace out the physical spins  intended, two adjacent rows
per spin. Hence one can compute physical properties such as
the entropy for the reduced density matrix.
In the Heisenberg picture the time evolution of the Majorana
fermions is
\begin{equation}
\frac{d}{dt}\,\check\gamma_k=\imi\left[\mathcal
  H,\check\gamma_k\right]=-2\imi\sum_iC_{ki}\check\gamma_i.
\end{equation}
If we denote the $k$th element of the $n$th eigenvector of $C$ as
$v_k^{(n)}$ and the corresponding eigenvalue as $\xi_k$, the
general solution of the above equation is $\check\gamma_k(t)=\mathds
T_{kl}(t)\check\gamma_l(0)$ with
\[\mathds T_{kl}(t)=\sum_n
v_k^{(n)}v_l^{(n)*}\expe^{-2\imi\xi_n t}.\]
Hence the correlation matrix evolves with time as
$\Gamma(t)=\mathds T(t)\Gamma(0)\mathds T^\transpose(t)$.
This is our equation of motion for the state defined by
$\Gamma_{ij}$. The time evolution is explicitly known, so that the
correlation 
matrix at any time $t$ is directly accessible and there is no
accumulation of errors with time.

\section{Entanglement evolution}
The state $\ket{\psi(t)}$ represented by the correlation matrix
$\Gamma(t)$, the entanglement of one part of the system
with the rest can be defined in terms of the von Neumann entropy,
$S=-\trace\rho\log_2\rho$ where
$\rho=\trace'\ket{\psi(t)}\bra{\psi(t)}$ amounts to tracing out some
part of the system, 
i.e. removing the $2(N-N')$ columns and rows corresponding to the
environment from $\Gamma(t)$. The reduced density matrix is
diagonal in the basis of the $\hat c$ fermions, and can be
written 
\begin{align}
  \rho=\prod_{k=1}^{N'}\frac1{1+e^{-\omega_k}}e^{-\omega_k\hat
    c^\dag_k\hat c_k}.
  \label{eq:rhodiag}
\end{align}
Here $\omega_k\in[0,\infty)$ are a set of parameters defining the
state. Given that 
the eigenvalues of $\Gamma(t)$ are $\pm\lambda_k$ with
$0<\lambda_k<1/2$, we find that  $\lambda_k=\frac12\tanh\frac12\omega_k$.
All states are thus defined by their $N'$ parameters $\omega_k$ or
$\lambda_k$. A pure state of the full system can also be defined by
the $2^N$ fermionic occupation numbers in the full $\hat c$ basis. 

The entropy of the reduced density matrix defined by the $\lambda_k$s
is now
\begin{equation}
  S=\sum_{k=1}^{N'}H\left(\frac12(1+2\lambda_k)\right)
\end{equation}
where $H(x)=-x\log_2x-(1-x)\log_2(1-x)$ is the binary entropy
function.
All states in the following are written in the basis of $\hat c$
fermions unless otherwise stated, that is the state is defined by a
series of $N$ binary digits, and we may for simplicity uniquely label
a state by its decimal number, e.g. $\ket{101}\equiv\ket{5_d}$.

\begin{figure}[htb]
  \onefigure[scale=1.1]{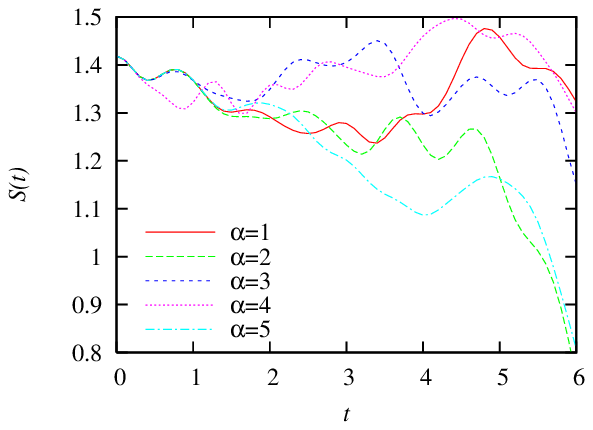}
  \caption{(Colour on-line) The entanglement entropy with time for a
    system of $N=10$ spins where 5 spins are traced
    out. The initial state in all cases is $\ket{\psi_0}=\ket{32_d}$.
    The impurity strength is $\delta=1$, placed at different
    positions $\alpha$ in the chain. Note that $\alpha$ and $N-\alpha+1$
    are equivalent positions.}
  \label{fig:S_N10}
\end{figure}

The simplest conceivable system is that of two particles, and with an
impurity of strength $\delta=1$ at one site, the entanglement entropy
evolves with a simple periodic behaviour. The periodicity
disappears for any larger system. Moreover, the specific choice of
impurity location has a profound effect on the evolution, as shown in
Fig. \ref{fig:S_N10}. Hence, any system larger than the trivial $N=2$
system is very irregular in its evolution.

\section{Thermal states}
Next we consider a state in thermal equilibrium with its
environment. Eventually, the time evolved state will be compared to
the closest thermal state.
A thermal state has density matrix 
\[\sigma=\frac1Z\expe^{-\beta\mathcal
  H}=\frac1Z\sum_n\expe^{-\beta E_n}\ket{\psi_n}\bra{\psi_n}\]
where $Z=\trace\expe^{-\beta\mathcal H}$ is the partition function,
$\beta$ is inverse 
temperature and $\ket{\psi_n}$ ($E_n$) are the eigenstates (-values)
of the Hamiltonian $\mathcal H$. To compare the thermal state $\sigma$
with some other state $\rho$, define the fidelity
$F(\sigma,\rho)=\trace\sqrt{\sigma^{1/2}\rho\sigma^{1/2}}$ between
these. The fidelity is one if the states are equal, and less
otherwise. Indeed, the fidelity is a distinguishability measure that
can be considered equivalent to a 
distance measure of mixed states \cite{Nielsen&Chuang}. However,
computing the 
fidelity between the mixed states is computationally hard unless the
density matrices commute. Hence, we will consider the classical
fidelity between the eigenvalue distributions of the two density
matrices, which is equal to the quantum fidelity above only in the
case where they commute,
\begin{equation}
  F_c(\sigma,\rho)
  =\prod_{k=1}^{N'}\frac{1+e^{-\frac12\left(\omega_k+\beta
      E_k\right)}}{\sqrt{\left(1+e^{-\omega_k}\right)\left(1+e^{-\beta E_k}\right)}}
  \label{fiddef}
\end{equation}
given that $E_k$ and $\omega_k$ both are ordered increasingly.

To compute the fidelity, we need a good estimate of the
inverse temperature $\beta$. To this end, consider a state in thermal
equilibrium in a canonical ensemble, which should obey the equality
\begin{equation}
  \ln Z=S_T-\beta U.
  \label{thermodynamicEOS}
\end{equation}
Here, $U=-\partial\ln Z/\partial\beta$ is the internal energy and
$S_T$ is the thermodynamical entropy. The entanglement entropy is
conventionally computed base two while the thermodynamic entropy is
computed in the natural base, hence $S_T=S\ln 2$. We find that
the equation of state (\ref{thermodynamicEOS}) can be fulfilled by
adjusting $\beta$ until equilibrium is reached. That is, with a guess
of $\beta$ and the energy spectrum of the Hamiltonian, one can find
$\ln Z$ in a neighborhood 
$\beta\pm\delta\beta$ and thus $U$. $S_T$ is given by the state
$\rho$, and by adjusting $\beta$ one can 
now find the intersection point where (\ref{thermodynamicEOS}) is
fulfilled. Coincidentally, this is 
very close to the temperature where the fidelity is highest, and
therefore adjusting to thermal equilibrium is closely equivalent to
doing a best fit to estimate $\beta$.

Given a reduced density matrix $\rho=\trace'\ket\psi\bra\psi$ of a 1D 
system such as a spin chain in state $\ket\psi$, this can become thermal
under time evolution in the sense that it has an eigenvalue
distribution which has fidelity close to unity with a thermal state of
some temperature $\beta^{-1}$. $\ket\psi$ cannot be an exact eigenstate of the
system's Hamiltonian since this would not give rise to a time
evolution. Therefore 
we apply the Hamiltonian (\ref{eq:H}) prepared in an initial state
$\ket{\psi_0}$, which is an eigenstate of the chain with
$\delta=0$. The initial state is chosen as a state where the binary 
digits defining the state in the $\hat c$ basis of the unperturbed
Hamiltonian are chosen at random.

\begin{figure}[htb]
  \twofigures[scale=1.1]{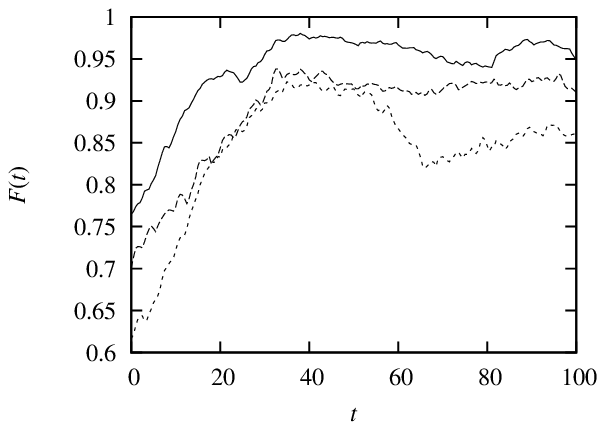}{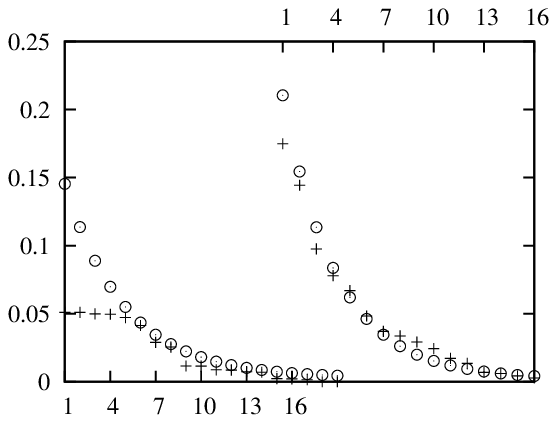}
  \caption{The fidelity of a state with time for three different
    initial states and $N=50$. The impurity is at the middle of the
    chain with strength $\delta=1$. At each time step a new estimate
    for the temperature of the state is applied. }
  \label{fig:fid_t}
  \caption{The eigenvalue distribution $e^{-\omega_k}$ of the reduced
    density matrix 
    for the state whose time evolution is drawn with full line in
    Fig. \ref{fig:fid_t} is plotted ($+$) along with the thermal
    distribution for a best fitted temperature ($\circ$), $e^{-\beta
      E_k}$. 
    To the left are the spectra at $t=0$ where $F=0.77$, while
    to the right is the spectra for the same state after time
    evolution at $t=30$ where $F=0.96$.}
  \label{fig:spect_t}
\end{figure}

We consider a system of $N=50$ spins, which allows us to investigate all
eigenvalues and compare to those of a thermal state. Also, for a
system of this size, the chance of a 
random state being close to a thermal state is negligible. In
particular, we find
that a randomly picked state on average has lower fidelity with a
thermal state as $N$ increases. In Fig.
\ref{fig:fid_t} the time evolution of the fidelity is shown for three
different initial states. We see that all states quickly gains a higher
fidelity, which is retained although there are some fluctuations. In
Fig. \ref{fig:spect_t} the eigenvalue distribution is plotted along
with the thermal spectrum for the best fitted temperature for two
values of $t$. We see clearly that the nature of the spectrum is
closer to the thermal spectrum after the initialization as reflected
in the fidelity. 

\begin{figure}[htb]
  \onefigure[scale=1.3]{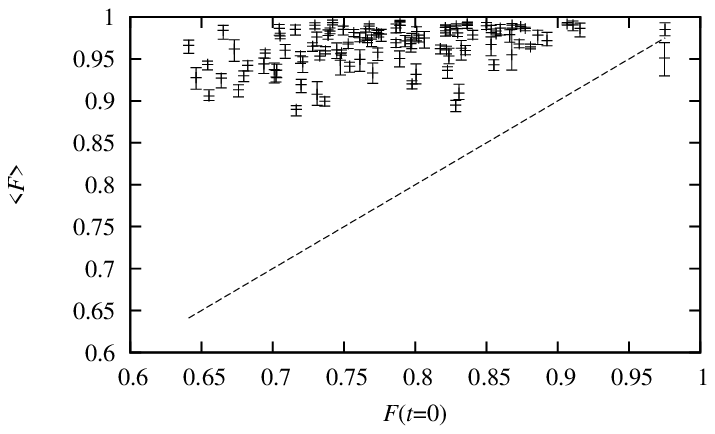}
  \caption{The fidelity of a random initial state with a thermal state
    on the
    $x$-axis and the fidelity of the state after a warmup period. The
    error bars indicate the range of the fidelity over the timespan
    recorded. Here the output state is recorded over the time
    $30<t<40$. The impurity is situated at the middle of the chain
    with strength $\delta=1$ and $N=50$. The straight line is where
    $\langle F\rangle=F(0)$, i.e. where there is no thermalization.}
  \label{fig:fidplot}
\end{figure}

To average out the random fluctuations, we define the average fidelity
recorded over some time span $t_1<t<t_2$, which we denote $\langle
F\rangle$. A new guess for $\beta$ is done at every time step. In Fig. 
\ref{fig:fidplot} the average fidelity after some initializing time is
plotted with 
the input fidelity for a selection of randomly picked
states. The output state is always more thermalized than the
input, in the figure indicated by the fact that all states lie
above the diagonal line, except where the input state by chance
already resembled a thermal state in its spectrum. However, there is
some variety as to how well the states thermalize.

\section{Entanglement induction}
\begin{figure}[htb]
  \twofigures[scale=1.1]{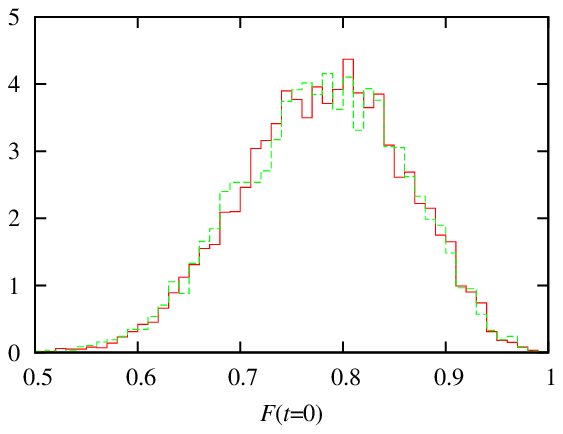}{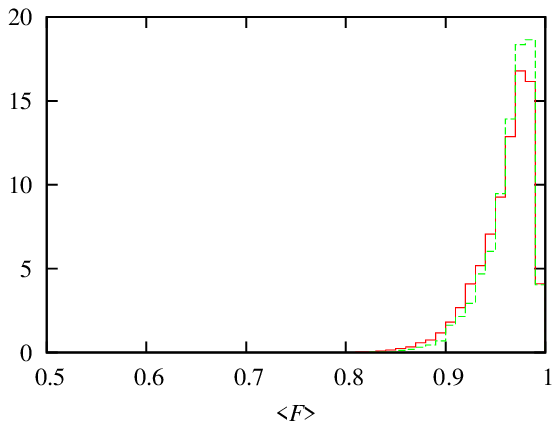}
  \caption{(Colour on-line) A histogram of the fidelity of a random state with 10,000
    samples and $N=50$. A magnetic impurity of strength $\delta=1$ is
    at mid-chain. The red line is for a normal chain, while the
    green line is when the chain is decoupled from its environment.}
  \label{fig:avg_thermalize1}
  \caption{(Colour on-line) The average fidelity after a time evolution. The fidelity
    is averaged over time $30<t<40$. The states that are evolved are
    the same as the initial states used in Fig. \ref{fig:avg_thermalize1}}
  \label{fig:avg_thermalize2}
\end{figure}

Finally, we investigate a different type of impurity, namely a
disconnection of the chain. To this end, consider the chain
defined by $\mathcal H'=\mathcal H-\sigma_{N'}^x\sigma_{N'+1}^x$.
The last term disconnects the chain described by the reduced
density matrix $\rho$ from its environment. Hence, there is no interaction
across the boundary, and the only communication between the two sides
is through the entanglement of the original state. It turns out that
the single impurity of cutting the bond does not induce a thermal
state in the sense that the output fidelity on average is no larger
than the input fidelity. However, reinstating the magnetic impurity in
addition to the decoupling of the chain makes the chain thermalize
again. This confirms that the magnetic impurity is necessary to
thermalize the state, though the physical coupling across the boundary
is not 
fundamentally important. As shown in Figs. \ref{fig:avg_thermalize1}
and \ref{fig:avg_thermalize2}, 
the output average fidelity is close to unity (i.e. $\langle F\rangle
> 0.9$) for most states 
even when the chain is decoupled, indeed the coupling does not seem to
make much difference as to whether the chain thermalizes or not.

Thus it is clear that the entanglement present in the state has enough
correlations to induce the thermalization of the state under this
dynamics. A state that is a product state,
$\ket\psi=\ket{\psi_{\mathrm{system}}}\otimes\ket{\psi_{\mathrm{environment}}}$
would remain a product state under Hamiltonians that do not provide
interaction between the system and the environment. However there 
is no clear connection between how entangled the state is, as
measured by the entanglement entropy, and how well it thermalizes.

\section{Conclusion}
We have seen that the reduced density matrix of a part of a spin chain
in a pure chain evolves with time such that the spectrum of the
reduced state has 
fidelity with that of a thermal state close to unity. Thus,
thermalization of a state can be achieved through simple unitary time
evolution. This even holds 
when the link between the subsystem and its environment is cut. Then
the induction of a thermal state must be 
caused by the entanglement correlations in the state.

The requirements on the total system is that it is
sufficiently large, with a non-zero energy per degree of freedom above
the vacuum energy, and that the unitary evolution is sufficiently
mixing. I.e. the wave function for the total system should not be
constrained by too many conservation laws to hamper it from visiting
typical regions of Hilbert space. Finally the subsystem under
consideration, which we describe by a density matrix by tracing out
all external degrees of freedom from the pure state, must perhaps be
small compared to the total system if we want to compare its density
matrix with one from a canonical or grand canonical ensemble. For this
to be the case it must be that {\em almost all} states in the total
Hilbert space, on the manifold satisfying the appropiate conservation
laws, will lead to a thermal state when tracing out the
external degrees of freedom. It can't be that the dynamics by miracle
drive the total system towards some special  states with this
property.

\acknowledgements
The author thanks Prof. K{\aa}re Olaussen for the ideas
behind this research. Also, Susanne Viefers and the NordForsk
network on {\it Low-dimensional physics: The theoretical basis of
  nanotechnology} is thanked for support and meetings.

\end{document}